\documentclass[conference]{IEEEtran}
\usepackage{array, rotating}
\usepackage{multirow, makecell}
\IEEEoverridecommandlockouts
\usepackage{cite}
\usepackage{amsmath,amssymb,amsfonts}
\usepackage{multirow}
\usepackage{multicol}
\usepackage{makecell}
\usepackage{algorithmic}
\usepackage{graphicx}
\usepackage{textcomp}
\usepackage{syntax}
\usepackage[final]{listings}
\usepackage{caption}
\usepackage{lipsum}
\usepackage{cuted}
\usepackage{xcolor}
\usepackage{textcomp}
\usepackage{tikz}
\usepackage{etoolbox}
\def\BibTeX{{\rm B\kern-.05em{\sc i\kern-.025em b}\kern-.08em
    T\kern-.1667em\lower.7ex\hbox{E}\kern-.125emX}}

\newcommand{\rating}[1]{%
  \begin{tikzpicture}[x=1ex,y=1ex]
    \begin{scope}
      \centering
      \clip (0,1) circle (1);
      \fill[black] (-1,0) rectangle (1,#1/50);
    \end{scope}
    \draw[black, thin, radius=1] (0,1) circle;
  \end{tikzpicture}%
}

\usepackage{listings}
\definecolor{gray}{RGB}{102,102,102}		
\definecolor{lightblue}{RGB}{0,102,153}		
\definecolor{lightgreen}{RGB}{102,153,0}	
\definecolor{bluegreen}{RGB}{51,153,126}	
\definecolor{magenta}{RGB}{217,74,122}		
\definecolor{orange}{RGB}{226,102,26}		
\definecolor{purple}{RGB}{125,71,147}		
\definecolor{green}{RGB}{113,138,98}		

\lstdefinelanguage{dreal}{
  morekeywords = [1]{set, logic, declare, fun, Int, Real, assert, check, sat, exit, and, delta},
  morekeywords = [2]{catch, do, for, if, else, switch, synchronized, while, try},
  morekeywords = [3]{width, height, pixelHight, displayHeight, displayWidth, focused, frameCount, frameRate, key, keyCode, keyPressed, mouseButton, mousePressed, mouseX, mouseY, pixels, pixelWidth, pmouseX, pmouseY},
  morekeywords = [4]{Array, ArrayList, Boolean, Byte, BufferedReader, Character, Class, Double, Float, Integer, HashMap, PrintWriter, String, StringBuffer, StringBuilder, Thread, boolean, byte, char, color, double, float, int, long, short, FloatDict, FloatList, IntDict, IntList, JSONArray, JSONObject, PFont, PGraphics, PImage, PShader, PShape, PVector, StringDict, StringList, Table, TableRow, XML},
  morekeywords = [5]{boolean(},
  keywordstyle = [5]\color{bluegreen},
  keywordstyle = [2]\color{lightgreen},
  keywordstyle = [3]\color{magenta},
  keywordstyle = [4]\color{orange},
  keywordstyle = [1]\color{blue},
  sensitive = true,
  morecomment = [l]{//},
  morecomment = [s]{/*}{*/},
  morecomment = [s]{/**}{*/},
  commentstyle = \color{gray},
  morestring = [b]",
  morestring = [b]',
  stringstyle = \color{purple}
}
\begin{document}

\setlength{\belowdisplayskip}{0pt} 

\title{On the Validation of Multi-Level Personalised Health Condition Model
}

\author{
\IEEEauthorblockN{Najma Taimoor}
\IEEEauthorblockA{\textit{Institute for Computer Technology} \\
\textit{Vienna University of Technology}\\
Vienna, Austria \\
e1155891@student.tuwien.ac.at}
\and
\IEEEauthorblockN{Semeen Rehman}
\IEEEauthorblockA{\textit{Institute for Computer Technology} \\
\textit{Vienna University of Technology}\\
Vienna, Austria \\
semeen.rehman@tuwien.ac.at}
}
\maketitle

\begin{abstract}
This paper presents a verification-based methodology to validate the model of personalised health conditions. The model identifies the values that may result in unsafe, unreachable, in-exhaustive, and overlapping states threaten otherwise threaten the patient’s life by producing false alarms by accepting suspicious behaviour of the target health condition. Contemporary approaches to validating a model employ various testing, simulation and model checking techniques to recognise such vulnerabilities. However, these approaches are neither systematic nor exhaustive and thus fail to identify those false values or computations that estimate the health condition at run-time based on the sensor or input data received from various IoT medical devices. We have demonstrated our validation methodology by validating our example multi-level model that describes three different scenarios of \emph{Diabetes} health conditions.
\end{abstract}


\section{Introduction} \label{sec:intro}
Model-based run-time monitoring of a health condition~\cite{Chase:2018} is a technique that monitors the health condition by comparing run-time estimation of the health condition (i.e., as determined by the input data received from sensors and other IoT medical devices with the expected estimation of the health condition (i.e., as formalised in the model). It raises an alarm when an inconsistency between the two estimations is detected and in case of an alarm, the monitor knows what exactly went wrong due to the violated description in the model. Since the monitor automatically monitors and controls the health condition, it is critically important to ensure that the model is valid because any false detection or evasion by the monitor may threaten the patient's health and eventually life. Typically, the model is valid, if it does not accept any false estimations of the health condition. Contemporary approaches to establishing the validity of the model employ different techniques. For instance, the approaches developed in~\cite{Cohen:2021, Van:2019} employ testing techniques to validate the model by running several tests and observing if any tests that accept false estimates. The other approaches~\cite{Esp:2021,Bongert:2019,Cheng:2014} employ simulation techniques to validate the model by executing various simulation runs based on different configurations of the model and by observing any runs that accept false estimations. However, testing and simulation-based approaches are not practical mainly because they are in-exhaustive and impractical on one hand (e.g., they fail to test/simulate all possible estimations due to required large computational resources and time), and are not systematic and rigorous on the other hand (e.g., there is no definition of a good test/simulation)~\cite{dij1976}. The authors in~\cite{Barbot:2015,Jones:2006,Boud:2014,Vatan:2020} works have employed verification techniques to validate the model, for example, by executing the model and checking if all possible runs of the formally specified model do not accept any false estimations. The verification techniques are rigorous as they produce a counterexample (i.e., traces of the run) exactly showing why the identified estimation is indeed false. However, the afore-mentioned model checking techniques can detect false estimations of the health condition if the model involves linear computations operating over regular integer values. However, in practice, a health condition is monitored by the values measured by various sensors and IoT-based medical devices which involve non-linear computations operating over real values with a variable error rate of the involved sensors. Furthermore, the data received from sensors and IoT devices may get compromised due to their limited resources and inherently unreliable data-sharing wireless communication channels.

We propose a novel approach that employs verification to validate a given model by identifying vulnerabilities in the model. The vulnerabilities results in false estimation of the health condition at run-time including
\begin{itemize}
        \item \emph{un-safe states:} the legitimate states (i.e., not breaching the model), but are unsafe for patients (i.e., compromise the underlying biological process),
        \item \emph{un-reachable states:} the states that are legitimate, but are not reachable due to underlying biological processes, and
        \item \emph{in-exhaustive and overlapping states}: the states are developed as a result of rules that describe different levels of the health condition, e.g., either the rules do not cover all possible levels, or the various rules describe the same level, respectively.
\end{itemize}
\begin{figure*}
    \centering
    \includegraphics[scale=0.60]{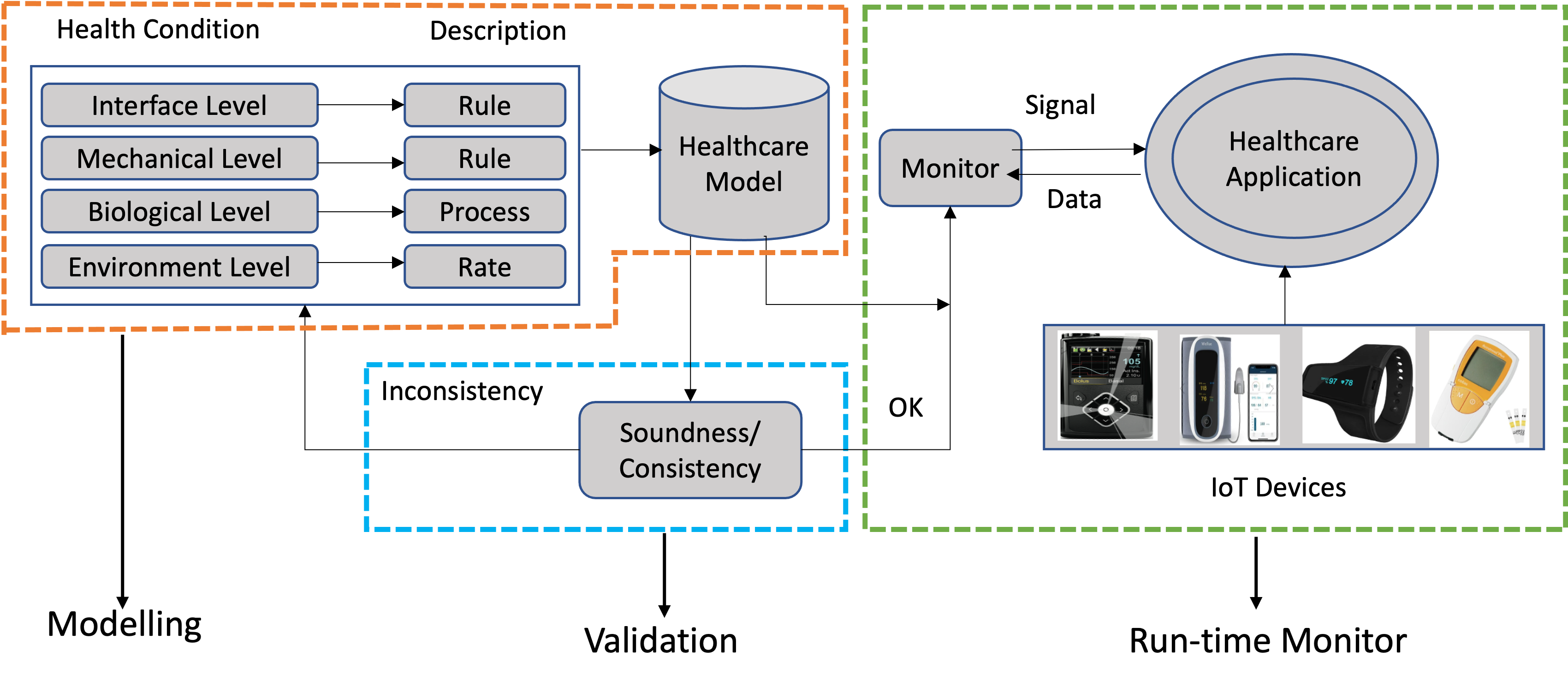}
    \caption{Monitoring of Personalised Healthcare Services}
    \label{fig:HM}
\end{figure*}
Our methodology detects the aforementioned vulnerabilities by producing a counterexample that includes a set of those values that constitute a false estimate. Once false estimations are identified, then either the model can be refined so that it does not accept those values or the set of values can be monitored at run-time to raise an alarm when detected. Our methodology is part of the larger project shown in Fig.~\ref{fig:HM}, whose aim is model-based monitoring of personalized health condition of a patient at run-time to ensure continuous operation in case of failure or attack, and patient's safety on one hand, and extending patient's life expectancy by optimising her multiple health conditions on the other hand. Fig.~\ref{fig:HM} shows the following key modules of our methodology:
\begin{enumerate}
     \item Modelling technique that allows multi-level modelling of a health condition with a fundamentally different abstraction that is a mix of fine-grained and coarse-grained on one hand, and integrates various modelling descriptions e.g., rules, process, rate and mathematical theories by overcoming their inherent limitations on the other hand, as explained in Section~\ref{sec:modelling} and~\cite{Najma2022a}.
     \item A methodology to validate the health condition model (developed in the previous step) ensuring that
     \begin{itemize}
         \item a given model adequately integrates various rules, processes, rates and mathematical theories that describe the health condition, and
         \item verifies that the given model only accepts clinically valid states of the health condition by identifying and eliminating un-reachable, un-safe and overlapping states as explained in Section~\ref{sec:validation}.
     \end{itemize}
     \item A methodology to develop a run-time monitor from the refined model (validated in the previous step). The monitor checks if a given input data from the application is consistent with the model and reports an error when it violates any model description. Specifically, the monitoring is free from false alarms because it not only reports the violation but also identifies the exact description and rule that was violated, which makes the monitoring more explainable and practical~\cite{Najma2022a}.
 \end{enumerate}
 
 The rest of the paper is organized as follows. Section~\ref{sec:relwork} presents related work for various techniques to model validation. Section~\ref{sec:modelling} explains our methodology that supports multi-level modeling of an example personalized health condition. Section~\ref{sec:validation} first introduces our methodology for validation of the model, then demonstrates the approach through three representative example scenarios. This manuscript is concluded in section~\ref{sec:conclusion}. 


\section{State-of-the-Art}\label{sec:relwork}
In this section, we present state-of-the-art techniques that are used to validate the models in general and personalised health condition models in particular, i.e., model testing, model simulation and model checking, respectively. As sketched in Table~\ref{table:sota}, we evaluate the aforementioned techniques against the following parameters:
\begin{itemize}
    \item \emph{validation} -- if the technique supports detection of invalid computations as well as data,
    \item \emph{completeness} -- if the technique detects invalid computations as well as data,
    \item \emph{diagnostic resolution} -- if the technique provides exact information of the detected invalid computation and data,
    \item \emph{multiple health condition} -- if the technique supports validation of multi-level model of personalized health conditions, and
    \item \emph{scalable} -- if the technique is scalable.
\end{itemize}

\begin{table*}
\begin{minipage}{0.8\textwidth}
\centering
\begin{tabular}{|*{8}{c|} }
 \hline
 \textbf{Technique} & \textbf{Method} & \multicolumn{2}{c|}{\textbf{Validation}} & \textbf{Completeness} & \textbf{Diagnostic Resolution} & \textbf{\makecell{Multiple \\health condition(s)}}  & \textbf{Scalable}\\ 
 
  \cline{3-4}\\
 &  & \textbf{Computing} & \textbf{Data} & & & &
 \\[0.5ex]
 \hline\hline
  \multirow{3}{*}{Model Testing} & Black-box~\cite{Walker:2019, Watson:2019} & \rating{25} & \rating{25} & \rating{20} & \rating{0} & \rating{0} & \rating{50}\\
     & White-box~\cite{Mezuk:2013} & \rating{25} & \rating{25} & \rating{20} & \rating{0} & \rating{0} & \rating{50}\\
     & Unit~\cite{Lucas:2020} & \rating{50} & \rating{25} & \rating{20} & \rating{0} & \rating{0} & \rating{50}\\
     & various AI~\cite{Kaushik:2020, Alfian:2018} & \rating{10} & \rating{10} & \rating{10} & \rating{0} & \rating{0} & \rating{100} \\\hline
     
     \multirow{3}{*}{Model Simulation} & Discrete event~\cite{Demir:2017} & \rating{50} & \rating{25} & \rating{20} & \rating{20} & \rating{0} & \rating{50}\\
     & Process~\cite{Khayal:2021} & \rating{50} & \rating{50} & \rating{20} & \rating{20} & \rating{0} & \rating{50}\\
     & Dynamic~\cite{Sluijs:2021} & \rating{50} & \rating{25} & \rating{20} & \rating{20} & \rating{0} & \rating{50}\\\hline
     
     \multirow{3}{*}{Model Checking} & Approximation~\cite{Herd:2013} & \rating{50} & \rating{0} & \rating{50} & \rating{100} & \rating{0} & \rating{50}\\
     & Symbolic~\cite{Mancini:2018} & \rating{100} & \rating{0} & \rating{50} & \rating{100} & \rating{0} & \rating{50}\\
     & \makecell{Numeric \& Symbolic \\(this work)} & \rating{100} & \rating{100} & \rating{100} & \rating{100} & \rating{100} & \rating{75}\\\hline
\end{tabular}
\caption{Validation of Personalized Health Condition Model}
\label{table:sota}
\end{minipage}
\end{table*}

\subsection{Model Testing}
In this approach, first various tests are generated from the model, and then the execution of the model or its implementation is observed to check if the execution passes any test. The tests essentially capture various incorrect estimates of the health condition. There are various testing approaches employed by the state of the art, e.g., black-box~\cite{Walker:2019, Watson:2019}, white-box~\cite{Mezuk:2013}, unit~\cite{Lucas:2020} and various AI~\cite{Kaushik:2020, Alfian:2018} techniques to test the execution. These approaches are limited mainly because they fail to test every possible estimation of the health condition due to the required large computational resources and time. Furthermore, as sketched in Table~\ref{table:sota}, these approaches partially support the detection of those computations that results in false estimation and have no support for detecting those sensors or input values that constitute a false estimation of the health condition. These approaches are scalable only for those models that involves the values with numerous permutations (e.g., Boolean, short int), and provide no diagnostic resolution (i.e., provide no information about what exactly went wrong when a false estimation test passes).

\subsection{Model Simulation}
In this approach, first various configurations are generated from the model, and then simulation of the model or its implementation are observed to check if any configuration leads to false estimation. The configurations essentially capture various incorrect estimates of the health condition. Such approaches employ different simulation techniques, e.g., discrete event ~\cite{Demir:2017}, process~\cite{Khayal:2021}, and dynamic ~\cite{Sluijs:2021} simulation to test the execution. These approaches are limited mainly because they fail to configure every possible estimation of the health condition. Furthermore, as sketched in Table~\ref{table:sota}, these approaches partially supports the detection of those computations and data that results in the false estimation of the health condition. Analogous to testing, these approaches are partially scalable and provide no diagnostic resolution.


\subsection{Model Checking}
In this approach, first health conditions and their estimates are formalized as a model, then the model is executed to automatically check if all possible runs of the model can detect any false estimate. The approach produces a counterexample when a false estimate is detected. The counterexample provides an exact execution trace of the detected false estimates. The estimates essentially model various constraints on input values of the health condition model. Such approaches employ different satisfiability techniques, e.g., approximation~\cite{Walker:2019, Watson:2019} and symbolic~\cite{Mezuk:2013} solving to check the model execution. These approaches are good at detecting only those computations that lead to a false estimation but have no support for detecting such values. Furthermore, as sketched in Table~\ref{table:sota}, these approaches do not support the validation of the multiple health condition model. However, they support the diagnostic resolution because as a result of the counterexample, such approaches exactly know what execution trace constituted the false estimation. These approaches are partially scalable due to the state explosion problem.

In comparison with the above-mentioned different approaches, our methodology follows a model checking approach. However, the working of our approach can not be directly compared with other approaches (testing and validation) due to the fundamentally different input and processing of the approaches. As shown in Table~\ref{table:sota}, our methodology combines numeric and symbolic decision procedures to detect not only those computations that lead to a false estimation but also can detect those values that constitute a false estimation. Furthermore, our methodology also supports detecting false estimations in a multi-level personalized health condition model. In case of a false estimation, our methodology produces a counterexample that includes a set of values (i.e., estimation of the health condition) that are legal but not real. The approach is even capable of detecting stealthy sensor or input values with a specific error rate using $\delta$-precision decision procedures~\cite{Gao:2013}.

\section{Multi-Level Modeling of Health condition}\label{sec:modelling}

Current approaches aim to personalize a single health condition with specific causes (e.g., genetics, inheritance) but are unable to personalize the overall health of a patient which requires an understanding of the biological relationship among different health conditions. We define personalised healthcare that takes into account the interdependent effect of multiple health conditions of patients at different levels namely \emph{interface} (observable parameters), \emph{mechanical} (delta error of bio-sensing techniques involved in IoTs), \emph{biological} (e.g., involved biological process, genes, proteins), and \emph{environmental} (exercise, diet, medication) that helps to extend the life expectancy of the patient by optimizing his/her multiple health conditions~\cite{Najma2022a}.

\begin{figure}[ht]
    \centering
    \includegraphics[scale=0.60]{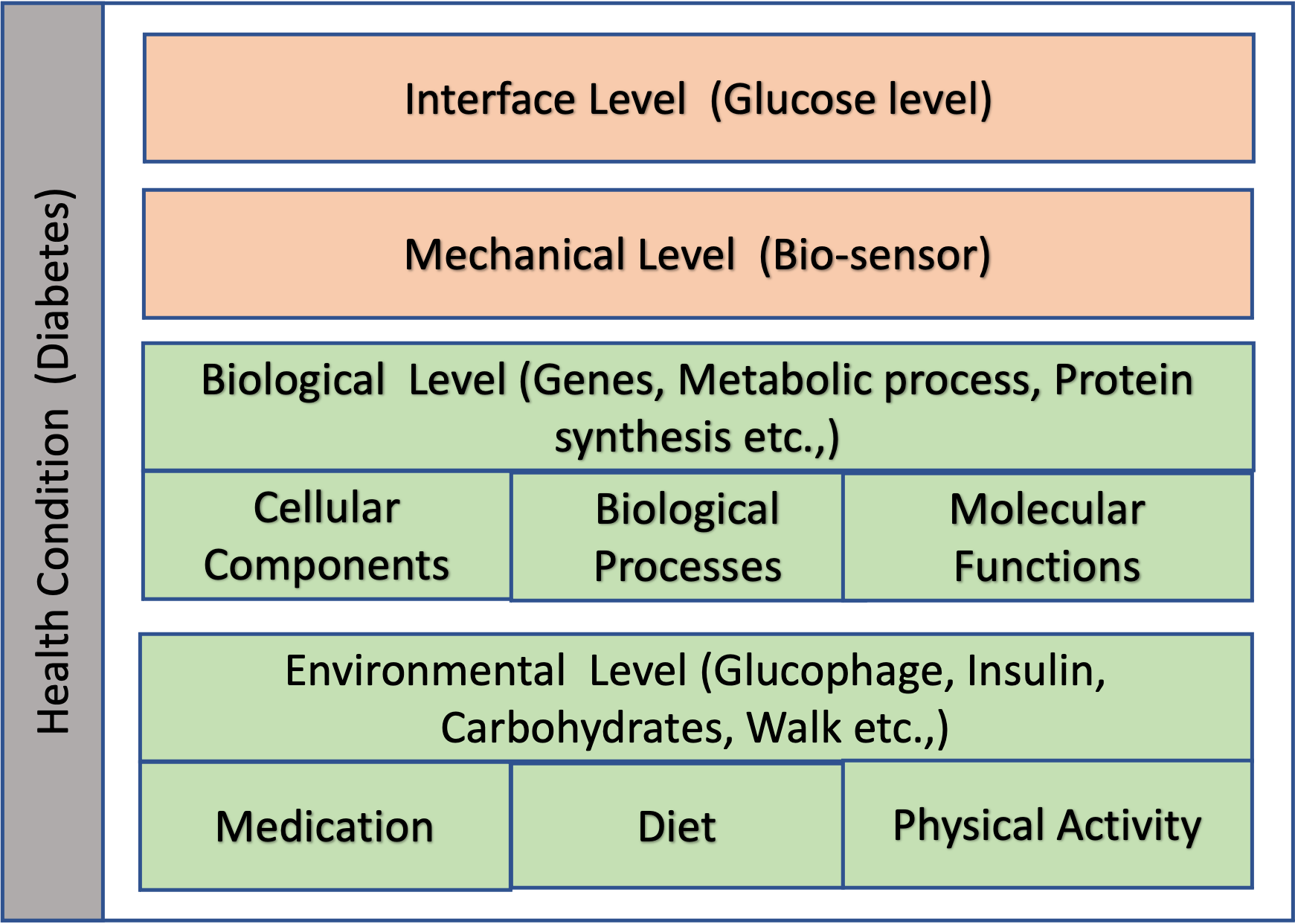}
    \caption{Multi-Level Health Condition Modelling Overview}
    \label{fig:HCO}
\end{figure}

 As a starting point, we have modelled a health condition using the modelling language that supports modelling a health condition as a logical relationship between different abstractions of the condition reflected at different levels. We model different abstractions (e.g., rules, processes and rates) as a logical relation where rule-based interface parameters are modelled as a Boolean function (i.e, predicate) over parameters with numeric and Boolean values. The incapability of the responsible biological process is modelled as a mathematical function that describes the rate of change (e.g., differential equation) in the biological process based on the interface parameters. The susceptibility of a health condition is also modelled as a mathematical function that describes how different external factors generate a susceptibility of the biological processes, and at the rate of its occurrence. For example, diabetes health condition (see Fig.~\ref{fig:HCO}) is modelled at the following levels.
 \begin{itemize}
    \item \emph{Interface level} models glucose concentration in the bloodstream (mmol/L) that respect certain ranges. 
    \item \emph{Mechanical level} model determines the inaccuracy of the interface level parameters that arise from the inherent unreliability of the underlying sensing techniques and IoT-based medical devices.
    \item \emph{Biological level} models incapacity of the biological process associated with the health conditions as determined by various elements of the responsible faulty genes, i.e., cellular component, biological process, and molecular function. For example, malfunctioning of the NFATc1 gene involved in diabetes causes a disorder Atrioventricular Septal Defect (AVSD) that causes cardiovascular and fetal genetic muscle diseases~\cite{genecard2010}.
    \item \emph{Environmental level} models susceptibility of the condition through external factors and activities, for example, medication (Glucophage) and lifestyle interventions (e.g., daily walk, carbohydrate-based diet)~\cite{magkos2020}.
 \end{itemize}
The former two levels describe discrete properties while the latter two describe continuous properties of the health condition. Though our modelling language is expressive enough to adequately model multiple health conditions and their interdependent effects, still we neither proved nor demonstrated that the language always allows the development of a valid model of the health condition, i.e., the model that does not accept any false abstractions of the health condition. Specifically, we show that the model is valid if and only if it is free from the vulnerabilities involving unsafe, unreachable, in-exhaustive, and overlapping states as explained in section~\ref{sec:intro}.

\begin{figure*}[ht]
    \centering
    \includegraphics[scale=0.55]{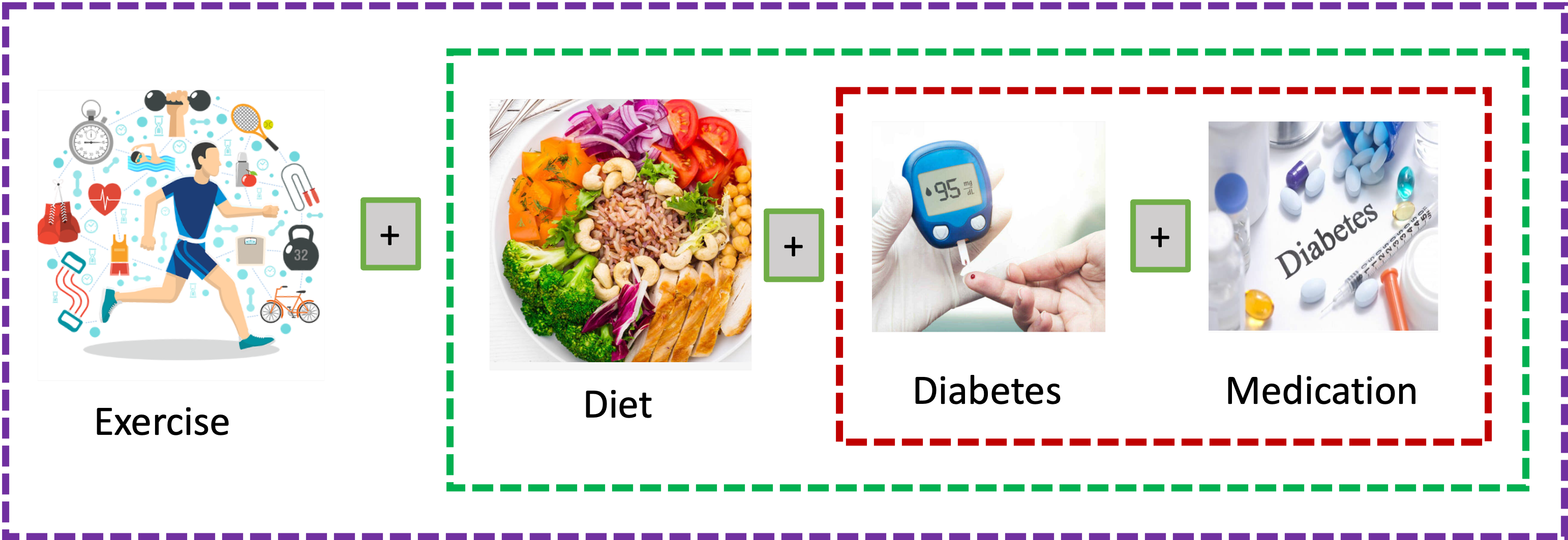}
    \caption{Validation of Diabetes Model Considering Medication, Diet, and Exercise }
    \label{fig:VM}
\end{figure*}


\begin{figure*}
\begin{equation}
    \label{eq:bg1}
    bg(f, gl, ge, a, p, l, map)=\begin{cases}
          (10 \leq gl+0.5 < 13.33) \quad &\text{if map = \{(gl, $\pm$0.5)\}, f is Yes, ge is F, 70 $<$ a $<$ 40,}
          \\\quad \wedge (10 \leq gl-0.5 < 13.33)  & \quad \text{p is Yes, l is Normal}\\
           \ldots
     \end{cases}
\end{equation}
\end{figure*}

In the next section, we introduce a methodology to validate the model of personalised health conditions by identifying the vulnerabilities (presented in Section~\ref{sec:intro}) in the model. In particular, the focus of this paper is to validate the model by showing that the model is free from unsafe conditions.

\section{Validation of the Model}\label{sec:validation}
The personalised health condition model presented in section~\ref{sec:modelling} adequately describes various fine-grained (e.g., biological and environmental level) and coarse-grained (e.g., interface and mechanical level) details of the condition at different levels. However, validation for the adequacy of our given example model (i.e., key information required to automatically verify the validity) has neither been established nor proved. Consequently, our given example model (discussed in section~\ref{sec:modelling}) is not safe and can threaten the patient's safety and life because of producing false alarms due to various vulnerabilities in the model arising from unsafe, unreachable, in-exhaustive, and overlapping states as explained in section~\ref{sec:intro}. 



Therefore, we present the methodology to validate the model of personalised health conditions by identifying various vulnerabilities in the model, e.g., the values that may result in the afore-mentioned unsafe, unreachable, in-exhaustive, or overlapping states. To identify such vulnerabilities, we use a non-linear SMT solver dReal~\cite{Gao:2013} that implements $\delta$-complete decision procedures by exploiting scalable numerical methods for handling non-linearity. A typical SMT solver~\cite{z3} is a tool for deciding the Boolean satisfiability of logical formulas in the background theories, e.g., arithmetic, bit-vectors, arrays, and uninterpreted functions. These solvers are used for extended static checking, predicate abstraction, test case generation, and bounded model checking over infinite domains. Analogously, the dReal solver takes a system model and desired properties of the model as an input and returns $unsat$ when the model does not satisfy the property or $\delta$-sat together with a set of values that all satisfy a $\delta$-perturbed form of the properties. We provide our model of the personalised health condition and ask the solver if it can find perturbed forms of the properties.

To demonstrate the effectiveness of our methodology, we simulate the following three different hypothetical but practical scenarios of diabetes health conditions based on the model equation~\ref{eq:bg1}. The scenarios are discussed later in the sub-sections and are sketched in Fig.~\ref{fig:VM} by red-dotted, blue-dotted and green-dotted rectangles, respectively.
\begin{enumerate}
    \item First scenario aims at validating the medication model such that \emph{the model does not accept any false estimation of the health condition when the patient is taking the required medicine as prescribed by the doctor}.
    
    \item Second scenario refines the first scenario through validating the additional diet model such that \emph{the model also does not accept any false estimation of the health condition when the patient is eating a health-condition-conscious diet as recommended by the doctor}.
    \item Third scenario refines the second scenario through validating the additional exercise model such that \emph{the model also does not accept any false estimation of the health condition when the patient is practising the exercise regularly as recommended by the doctor}.
   
\end{enumerate}
For demonstration, we simply show that the model is valid if and only if it is free from unsafe states. Our methodology assures that the model can detect accidentally or intentionally compromised input values which do not breach the model but do not provide the correct estimation of the health condition. Consequently, the approach assures that the multi-level model always correctly estimates the health condition.
\vspace{-0.5cm}
\subsection{Validation of Medication Model}\label{subsec:vmm}
This scenario includes a model where the diabetes health condition of a patient and the effect of the medication are modelled.
The goal is to validate the model by identifying susceptible values even when the health condition is in a normal state, i.e., diabetes level is normal and the patient is regularly taking the required medicine. For demonstration and reader's convenience, we simplify the model of the scenario which is presented in Equation~\ref{eq:bg1} with the following assumptions:
\begin{itemize}
    \item the state of diabetes is determined by a sensor that measures blood glucose value in the patient's body,
    \item a medicine that controls diabetes through injecting insulin unit in the patient's body is also determined by the sensor,
    \item the system runs in a discrete mode where the afore-mentioned values change in a discrete-time interval, and
     \item the sensor values received at earlier interval (e.g., at time $t$) are trusted. 
\end{itemize}

Considering the above assumptions, a health condition of diabetes can be modelled as 
\begin{equation}\label{eq:modeq}
    BG(t+1)\footnote{The variable BG(t+1) is represented as bg@t1 in dReal as shown in Listing~\ref{lst:ex1}. All the model variables are encoded in the dReal following the same analogy as shown in the corresponding listings.} = BG(t) - (I(t+1) - \delta)
\end{equation}
which says that the value of blood glucose (BG) at time $t+1$ equals the value of blood glucose at previous time $t$ minus the insulin units (I) taken at time $t+1$ and the injection error $\delta$. For simplicity, we assume that the injection of insulin units instantly reduces the blood glucose in the body.

For identifying the susceptible, let us simulate the following hypothetical scenario: if a blood glucose at time $t$ is $BG(t)=14$ and insulin units injected at time $t+1$ is $I(t+1)=1.5$ that results in a new blood glucose $BG(t+1)=12$ at time $t+1$ with $\delta$ error of $-0.5$. If the sensors provide secure and correct values $BG(t+1)=12$, $BG(t)=14$, and $I(t+1)=1.5$ to the model, they will be accepted validating the model because the values satisfy the model represented in Equation~\ref{eq:modeq} as $12 = 14 - (1.5 + 0.5)$. However, due to inherently insecure and unreliable communication of sensor network, or adversary's interference, the sensor values at run-time may be hoaxed
\begin{enumerate}
    \item either by compromising value $BG(t+1)=13$, which will be rejected by the model because the value doesn't satisfy the model as represented by equation~\ref{eq:modeq}  because $13 \ne 14 - (1.5 + 0.5)$,
    \item or by compromising values $I(t+1)=2.5$ and $\delta=+0.5$, which will be accepted by the model satisfying Equation~\ref{eq:modeq} because $12 = 14 - (2.5 - 0.5)$. Evidently these values are a result of accidental (e.g., unreliable sensor resources) or intentional (e.g., adversarial) incident.
\end{enumerate}
Our methodology can identify the latter compromised values using the solver by providing the aforementioned model as an input (see Listing~\ref{lst:ex1}) that produced the results (see Listing~\ref{lst:res1}) successfully identifying the compromised values that satisfied the model but are not real. Practically, there could be several reasons that may result in such susceptible values, e.g.,
\begin{enumerate}
    \item either the sensor or medical device is producing incorrect values due to malfunctioning of their software or hardware,
    \item or some adversary may have compromised the values,
    \item or the values indicate accurate but unprecedented and exceptional health condition of the patient due to an unknown factor (e.g., inappropriate diet, short of exercise or underlying other health condition).
\end{enumerate}

\begin{lstlisting}[language=dreal, caption=Example Model, label={lst:ex1}]
(set-logic QF_NRA)
(declare-fun bg@t () Int)
(declare-fun bg@t1 () Int)
(declare-fun i@t1 () Real)
(declare-fun err () Real)
(assert (and (= bg@t1 12) (= bg@t 14)))
(assert (= err 0.5))
(assert (and (<= 1 i@t1) (<= i@t1 5)))
(assert (= bg@t1 (- bg@t (- i@t1 err))))
(check-sat)
(exit)
\end{lstlisting}

Once the suspicious values are detected in the former two cases, either the model can be refined to restrict such values or can be handled at run-time. While in the last case, a practitioner can further investigate the health condition by carrying advance tests of the underlying condition to determine the actual state of the health. To refine the model, we validate the adequacy of the model for unknown factors, e.g., the effect of inappropriate diet on the health condition. In the next subsection, we first define the model to incorporate the unknown factor and discuss the validity of the refined model.


\begin{lstlisting}[language=dreal, caption=Identified Invalid Values, label={lst:res1}]
delta-sat with delta = 0.01
bg@t : [14, 14]
bg@t1 : [12, 12]
i@t1 : [2.5, 2.5]
err : [0.5, 0.5]
\end{lstlisting}

\subsection{Validation of Medication and Diet Model}\label{subsec:vmdm}
This scenario includes a model where the diabetes health condition of a patient and the effect of medication and diet (i.e., carbs) are modelled.
The goal is to validate the model by identifying susceptible values even when the health condition is in a normal state, i.e., diabetes level is normal, the patient is regularly taking the required medicine and the desired diet. For demonstration, we simplify and refine the model of the scenario which is presented in Equation~\ref{eq:bg1} with the following assumptions:
\begin{itemize}
    \item the state of diabetes is determined by a sensor that measures blood glucose value in the patient's body,
    \item a medicine that controls diabetes through injecting insulin in the patient's body is also determined by the sensor,
    \item a diet intake ($D_{in}$) that adds a specific amount of carbohydrates ($C_{in}$) in the patient's body which is monitored by the sensor,
    \item the system runs in a discrete mode where the afore-mentioned values change in a discrete-time interval.
\end{itemize}

Considering the above assumptions, a health condition of diabetes can be modelled as 

\begin{equation}\label{eq:modeq1}
\begin{aligned}
    BG(t+1) = {} & BG(t) & {} \\
              & + (D_{in}(t+1) \times C_{in}(t+1)) & \\
              & - (I(t+1) - \delta)
\end{aligned}
\end{equation}
which says that the value of blood glucose at time $t+1$ equals
\begin{enumerate}
    \item the value of blood glucose at previous time $t$
    \item plus the diet intake multiplied by the carbs in the diet, which  computes collective affect of carbs on the health condition and
    \item subtract the insulin units taken at time $t+1$ and the injection error $\delta$. 
\end{enumerate} 

For simplicity, we assume that insulin injection instantly reduces the blood glucose in the body.

For identifying the susceptible, let us simulate the following hypothetical scenario: if a blood glucose at time $t$ is $BG(t)=14$ and insulin units injected is $I(t+1)=2.5$ that results in a new blood glucose $BG(t+1)=12$ at time $t+1$ with $\delta$ error of $-0.5$. Furthermore, the diet intake at time is $D_{in}=2$ at $t+1$ that contains $50\%$ carbohydrates $C_{in}=0.5$. If the sensors provide secure and correct values $BG(t+1)=12$, $BG(t)=14$, $I(t+1)=2.5$, $D_{in}=2$ and $C_{in}=0.5$ to the model, they will be accepted validating the model because the values satisfy the model represented in Equation~\ref{eq:modeq1} as $12 = 14 + (2\times 0.5) - (2.5 + 0.5)$. However, due to inherently insecure and unreliable communication of sensor network, or adversary's interference, the sensor values at run-time may be hoaxed
\begin{enumerate}
    \item either by compromising value $BG(t+1)=13$, which will be rejected by the model because the value don't satisfy the model as represented by Equation~\ref{eq:modeq1}  because $13 \ne 14 + (2\times 0.5) - (2.5 + 0.5)$,
    \item or by compromising values $D_{in}=1$, $I(t+1)=3.0$ and $\delta=+0.5$, which will be accepted by the model satisfying Equation~\ref{eq:modeq1} because $12 = 14 + (1\times 0.5) - (3.0 - 0.5)$. Evidently these values are a result of accidental (e.g., unreliable sensor resources or medical devices) or intentional (e.g., adversarial) incident.
\end{enumerate}


Analogous to the previous scenario, our methodology can again identify the latter compromised values using the solver by providing the afore-mentioned model as an input (see Listing~\ref{lst:sc2}) that produced the results (see Listing~\ref{lst:res2}) successfully identifying the compromised values that satisfied the model but are not real. The susceptible values are result of the reasons as discussed in the section~\ref{subsec:vmm}. 



\begin{lstlisting}[language=dreal, caption=Example Model, label={lst:sc2}]
(set-logic QF_NRA)
(declare-fun bg@t () Int)
(declare-fun bg@t1 () Int)
(declare-fun i@t1 () Real)
(declare-fun diet_in@t1 () Real)
(declare-fun carbs_in@t1 () Real)
(declare-fun err () Real)
(assert (and (= bg@t1 12) (= bg@t 14)))
(assert (and (<= -0.5 err) (<= err 0.5)))
(assert (= err 0.5))
(assert (= carbs_in@t1 0.5))
(assert (and (<= 0 i@t1) (<= i@t1 3)))
(assert (and (<= 0 diet_in@t1) (<= diet_in@t1 3)))
(assert (and (<= 0 carbs_in@t1) (<= carbs_in@t1 3)))
(assert (= bg@t1 (- (+ bg@t (* diet_in@t1 carbs_in@t1)) (- i@t1 err))))
(check-sat)
(exit)
\end{lstlisting}

To handle the suspicious values,
we refine and validate the adequacy of the model for other unknown factors, e.g., the effect of inappropriate exercise on the health condition. In the next subsection, we further refine the model to incorporate the unknown factor and discuss the validity of the refined model.

\begin{lstlisting}[language=dreal, caption=Identified Invalid Values, label={lst:res2}]
delta-sat with delta = 1
bg@t : [14, 14]
bg@t1 : [12, 12]
i@t1 : [2.5, 3]
diet_in@t1 : [0, 1]
carbs_in@t1 : [0.5, 0.5]
err : [0.5, 0.5]
\end{lstlisting}

\subsection{Validation of Medication, Diet, and Exercise Model}\label{subsec:vmdem}
This scenario includes a model where the diabetes health condition of a patient, the effect of medication, diet (i.e., carbs) and exercise are modelled.

The goal is to validate the model by identifying susceptible values even when the health condition is in a normal state, i.e., diabetes level is normal, the patient is regularly taking the required medicine, desired diet and practising appropriate exercise. For demonstration and reader's convenience, we simplify model of the scenario which is presented in Equation~\ref{eq:bg1} with the following assumptions:
\begin{itemize}
    \item the state of diabetes is determined by a sensor that measures blood glucose value in the patient's body,
    \item a medicine that controls diabetes through injecting insulin in the patient's body is also determined by the sensor,
    \item a diet intake that adds a specific amount of carbohydrates in the patient's body which is monitored by the sensor,
    \item an exercise (EX) helps to reduce the effect of diet (i.e., carbohydrates) in the patient's body which is also monitored by the sensor,
    \item the system runs in a discrete mode where the afore-mentioned values change in a discrete-time interval.
\end{itemize}


Considering the above assumptions, a health condition of diabetes can be modelled as 

\begin{equation}\label{eq:modeq2}
\begin{aligned}
    BG(t+1) = {} & BG(t) & {} \\
              & + \frac{D_{in}(t+1) \times C_{in}(t+1)}{EX(t+1)} & \\
              & - (I(t+1) - \delta)
\end{aligned}
\end{equation}
which says that the value of blood glucose at time $t+1$ equals
\begin{enumerate}
    \item the value of blood glucose $BG$ at previous time $t$
    \item plus the diet intake $D_{in}$ multiplied by the carbs $C_{in}$ in the diet whose affect is reduced by the factor of exercise $EX$, which collectively computes the affect of exercise and diet on the health condition and
    \item subtract the insulin $I$ taken at time $t+1$ from the injection error $\delta$. 
\end{enumerate} 
For simplicity, we assume that the injection of insulin units instantly reduces the blood glucose in the body.

For identifying the susceptible, let us simulate the following hypothetical scenario: if a blood glucose at time $t$ is $BG(t)=14$ and insulin units injected at time $t+1$ is $I(t+1)=2.5$ that results in a new blood glucose $BG(t+1)=12$ at time $t+1$ with $\delta$ error of $-0.5$. While the diet intake at time is $D_{in}=1$ at $t+1$ that contains $50\%$ carbohydrates $C_{in}=0.5$ whose affect is reduced by the factor of exercise performed $EX(t+1)=0.5$. If the sensors provide secure and correct values $BG(t+1)=12$, $BG(t)=14$, $I(t+1)=2.5$, $D_{in}=1$ and $C_{in}=0.5$ to the model, they will be accepted validating the model because the values satisfy the model represented in Equation~\ref{eq:modeq2} as $12 = 14 + \frac{1\times 0.5}{0.5} - (2.5 + 0.5)$. However, due to inherently insecure and unreliable communication of sensor network, or adversary's interference, the sensor values at run-time may be compromised
\begin{enumerate}
    \item either by compromising value $BG(t+1)=13$, which will be rejected by the model because the value don't satisfy the model as represented by Equation~\ref{eq:modeq1}  because $13 \ne 14 + \frac{1\times 0.5}{0.5} - (2.5 + 0.5)$,
    \item or by compromising values $I(t+1)=2.75$, $D_{in}(t+1)=0.25$ and $\delta=+0.5$, which will be accepted by the model satisfying Equation~\ref{eq:modeq2} because $12 = 14 + \frac{0.25\times 0.5}{0.5} - (2.75 - 0.5)$. Evidently these values are a result of accidental (e.g., unreliable sensor resources or medical devices) or intentional (e.g., adversarial) incident.
\end{enumerate}
Analogous to the previous scenario, our methodology can again identify the latter compromised values using the solver by providing the afore-mentioned model as an input (see Listing~\ref{lst:sc3}) that produced the results (see Listing~\ref{lst:res3}) successfully identifying the compromised values that satisfied the model but are not real. The susceptible values are result of the reasons as discussed in the section~\ref{subsec:vmdm}. 

\begin{lstlisting}[language=dreal, caption=Example Model, label={lst:sc3}]
(set-logic QF_NRA)
(declare-fun bg@t () Int)
(declare-fun bg@t1 () Int)
(declare-fun i@t1 () Real)
(declare-fun diet_in@t1 () Real)
(declare-fun carbs_in@t1 () Real)
(declare-fun ex@t1 () Real)
(declare-fun err () Real)
(assert (and (= bg@t1 12) (= bg@t 14)))
(assert (and (<= -0.5 err) (<= err 0.5)))
(assert (= err 0.5))
(assert (and (<= 0 ex@t1) (<= ex@t1 0.5)))
(assert (= carbs_in@t1 0.5))
(assert (and (<= 0 i@t1) (<= i@t1 3)))
(assert (and (<= 0 diet_in@t1) (<= diet_in@t1 3)))
(assert (and (<= 0 carbs_in@t1) (<= carbs_in@t1 3)))
(assert (= bg@t1 (- (+ bg@t (/ (* diet_in@t1 carbs_in@t1) ex@t1)) (- i@t1 err))))
(check-sat)
(exit)
\end{lstlisting}

Analogous to the previous scenario, once the suspicious values are detected in the former two cases, either the model can be refined to restrict such values or can be handled at run-time. While in the last case, a practitioner can further investigate the health condition by carrying advance tests of the underlying condition to determine the actual health state.

\begin{lstlisting}[language=dreal, caption=Identified Invalid Values, label={lst:res3}]
delta-sat with delta = 1
bg@t : [14, 14]
bg@t1 : [12, 12]
i@t1 : [2.5, 2.75]
diet_in@t1 : [0, 0.25]
carbs_in@t1 : [0.5, 0.5]
ex@t1 : [0.25, 0.5]
err : [0.5, 0.5]
\end{lstlisting}


\section{Conclusion and Future Work}\label{sec:conclusion}

We have presented a methodology to validate the multi-level health condition model. The effectiveness of the methodology is successfully demonstrated by validating three different scenarios of the Diabetes health condition by detecting those values that were accepted by the model but constituted a false estimate of the health condition. As a next step, we are validating the complete model. Later, we will develop a run-time monitor from the validated model to check the consistency of the health condition estimation derived from the model with the real-time estimation of the health condition as determined by the sensors and IoT-based medical devices. When an inconsistency is detected, the monitor reports an error and identifies the exact violated model description which will make the monitoring explainable. Furthermore, our methodology will help the practitioners in improving the management of the health condition by identifying its unsafe, unreachable, in-exhaustive, and overlapping states.

\bibliographystyle{IEEEtran}
\bibliography{paper}
%



\vspace{12pt}

\end{document}